\documentclass[preprint,11pt]{elsarticle}

\usepackage{lineno,hyperref}
\usepackage{amssymb, amsmath,bm}
\usepackage{graphicx}
\usepackage{graphics}
\usepackage{latexsym}
\usepackage{booktabs}
\usepackage[utf8]{inputenc}
\usepackage{fullpage}
\journal{Physics Letters B}

\begin{document}

\begin{frontmatter}

\title{The first large-scale shell-model calculation of the two-neutrino double beta decay of $^{76}$Ge to the excited states in $^{76}$Se}

\author{J.~Kostensalo}\address{Natural Resources Institute Finland, Yliopistokatu 6B, FI-80100 Joensuu, Finland}%

\author{J.~Suhonen}
\address{Department of Physics, University of Jyv\"askyl\"a, P.O. Box 35, FI-40014, Jyv\"askyl\"a, Finland}

\author{K.~Zuber}
\address{Institut f\"ur Kern- und Teilchenphysik, Technische Universität Dresden,
Zellescher Weg 19, 01069 Dresden, Germany}

\begin{abstract}
Large-scale shell-model calculations were carried out for the half-lives and branching ratios of the $2\nu\beta\beta$ decay of $^{76}$Ge to the ground state and the lowest three excited states $2_1^+$, $0_2^+$ and $2_2^+$ in $^{76}$Se. In total, the wave functions of more than 10,000 intermediate $1^+$ states in $^{76}$As were calculated in a three-step procedure allowing an efficient use of the available computer resources. In the first step, 250 lowest states, below some 5 MeV of excitation energy, were calculated without truncations within a full major shell $0f_{5/2}-1p-0g_{9/2}$ for both protons and neutrons. The wave functions of the rest of the states, up to some 30 MeV, were computed in two more steps by introducing two consecutive stages of truncation. The computed magnitudes of the $2\nu\beta\beta$ nuclear matrix elements (including the value of the axial-vector coupling $g_{\rm A}$), $\vert M_{2\nu}\vert g_{\rm A}^2$, converged to the values 0.168$g_{\rm A}^2$, $1.2\times10^{-3}$$g_{\rm A}^2$, 0.121$g_{\rm A}^2$, and $3.1\times10^{-3}$$g_{\rm A}^2$ for the $0^+_{\rm g.s.}$, $2^+_1$, $0^+_2$, and $2^+_2$ states, respectively. Using up-to-date phase-space integrals, the corresponding branching ratios were derived to be 99.926\%, 4.4$\times10^{-5}$\%, 0.074\% and 2.5$\times10^{-7}$\%. The experimental half-life $(1.926\pm0.094)\times10^{21}$ yr of the ground-state transition was used to derive the value $g_{\rm A}=0.80\pm0.01$ for the axial-vector coupling, which is consistent with other shell-model calculations suggesting a quenched value of $g_{\rm A}$. Using this value of $g_{\rm A}$, predictions for the transition half-lives were derived.
\end{abstract}


\end{frontmatter}



Undoubtedly, the nuclear double beta decay is one of the most relevant issues in today's neutrino and nuclear-structure physics. Experimentally, the most interesting nuclei decay by the double-$\beta^-$ mode, i.e., by two-neutrino double beta ($2\nu\beta^-\beta^-$) and neutrinoless double beta ($0\nu\beta^-\beta^-$) modes, hereafter simply denoted as $2\nu\beta\beta$ and $0\nu\beta\beta$ decays. The $0\nu\beta\beta$ mode is the more interesting one owing to its connections to the beyond-the-standard-model physics, with implications to lepton-number violation, neutrino masses and Majorana nature of the neutrino. The $2\nu\beta\beta$ mode is allowed in the standard model and the related half-lives have thus far been measured for more than ten nuclei \cite{Barabash2020}. Although the $2\nu\beta\beta$ mode is not as interesting in terms of beyond-the-standard-model physics, it is interesting in terms of nuclear structure \cite{Suhonen1998}, in terms of accessing the effective value of the weak axial coupling $g_{\rm A}$ \cite{Suhonen2017,Suhonen2019}, and in terms of being an irreducible background component in $0\nu\beta\beta$ measurements \cite{Ejiri2019}. In the history of double-beta-decay measurements and calculations $^{76}$Ge, decaying to the ground state and excited states of $^{76}$Se, has been one of the flagship cases. In this Letter, we present the first large-scale shell-model calculation of the nuclear matrix elements (NMEs) of the $2\nu\beta\beta$ decay of $^{76}$Ge to the $0^+$ ground state ($0^+_{\rm g.s.}$) and $2^+_1$, $0^+_2$ and $2^+_2$ excited states in $^{76}$Se.\\
Experimentally, the double-beta-decay search for $^{76}$Ge is performed with HPGe semiconductor detectors as suggested by Fiorini \textit{et al.} \cite{Fiorini1967}. Due to the superb energy resolution of such detectors, it is convenient to use them for searches of peak-like spectra, like is the case for the $0\nu\beta\beta$ decay of $^{76}$Ge. The two dominant double-beta-decay modes of $^{76}$Ge are the $2\nu\beta\beta$ decay with the emission of two electrons providing a continuous energy spectrum while the neutrinoless decay mode results in a peak at the $Q$ value. The $Q$ value has been measured to be 2039.061(7) keV \cite{Mount2010}. 
To observe such rare event as double beta decays requires a very low background and various components contribute. Among them are the installation of detectors deep underground, material screening to deduce the background level especially from the natural decay chains of U and Th. Isotopic enrichment in $^{76}$Ge is nowadays mandatory. Further techniques like pulse-shape analysis and veto systems against muons help to  reduce background further. The driving
experiments in the last decade have been the MAJORANA demonstrator and the GERDA experiment, employing bare HPGe detectors, the latter within liquid argon (see the review \cite{Avignone2019}). The GERDA has recently measured the $2\nu\beta\beta$-decay half-life of $^{76}$Ge to the ground state of $^{76}$Se to be
$T_{1/2}^{\beta\beta}=(1.926\pm 0.094)\times 10^{21}$ yr \cite{GERDA2015}.

%

On the nuclear-structure side, the $2\nu\beta\beta$ decay of $^{76}$Ge to the ground and excited states has attracted attention mainly in the nuclear-structure community specialized to many-body frameworks based on the quasiparticle random-phase approximation (QRPA) \cite{Suhonen2007}. The ground-state transition has usually been dealt with by the use of the proton-neutron QRPA (pnQRPA) \cite{Suhonen2012} and the excited-state transitions have been handled by the use of various extensions of the QRPA and pnQRPA \cite{Schuck2021}. In Civitarese \textit{et al.} \cite{Civitarese1994} foundations for the multiple commutator model (MCM) were laid, and further application of the theory to many $2\nu\beta\beta$ decays was performed in Aunola \textit{et al.} \cite{Aunola1996}. Various higher-QRPA frameworks were applied to $2\nu\beta\beta$ decays by Bobyk \textit{et al.} \cite{Bobyk1995}, Stoica \textit{et al.} \cite{Stoica1996} and Raduta \textit{et al.} \cite{Raduta2007}. After the introduction of the renormalised QRPA (RQRPA) by Toivanen \textit{et al.} \cite{Toivanen1995}, the theory was applied to $2\nu\beta\beta$ decays by the same authors in \cite{Toivanen1997} and by Schwieger \textit{et al.} in \cite{Schwieger1998}. No shell-model calculations before the present one could be achieved due to limited computational resources which could not match the number of active valence particles in a realistic valence space covering the $0f_{5/2}-1p-0g_{9/2}$ major shell, for both protons and neutrons. In particular, the involved nuclei, $^{76}$Ge, $^{76}$Se and the intermediate nucleus $^{76}$As, are almost in the middle of the $0f_{5/2}-1p-0g_{9/2}$ major shell in terms of the proton ($Z=32-34$) and neutron ($N=42-44$) numbers.


The half-life of $2\nu\beta\beta$ decay is given by \cite{Suhonen1998}
\begin{equation}
t^{(2\nu)}_{1/2} = \frac{1}{G^{(2\nu)}g_{\rm A}^4|M_{2\nu}|^2},
\label{eq:2vbbhl}
\end{equation}
where $G^{(2\nu)}$ is the phase-space integral which captures the kinematics of the decay. For the evaluation of the phase-space integral we use the values (readily available for the decays to $0^+_{\rm g.s.}$, $0^+_2$, and $2^+_1$) and expressions given in \cite{Neacsu2016}. The quantity $M_{2\nu}$ is the $2\nu\beta\beta$ NME which encodes the nuclear-structure information and can be written for $\beta^-\beta^-$ decay from the initial $0^+$ ground state ($0^+_{\rm g.s.}$) as
\begin{equation}
M_{2\nu} = \sum_m \frac{(J^+|| \sigma\tau^- || 1^+_m)
( 1^+_m|| \sigma\tau^- || 0^+_{\rm g.s.} )}{\sqrt{J+1}([\frac{1}{2}Q_{\beta\beta} + E(1^+_m)-M_i]/m_e +1)^{k}},
\end{equation}
where $J$ is the spin of the final state (0 or 2 in the present case), $m_e$ is the electron rest mass, $E(1^+_m)-M_i$ is the energy difference between 
the $m^{\rm th}$ intermediate 1$^+$ state and the ground state of the initial nucleus and 
$Q_{\beta\beta}$ is the energy released in the decay, i.e., the $Q$ value. The power $k$ assumes the value $k=1$ for $J=0$ and $k=3$ for $J=2$ (see the derivation and discussion in \cite{Doi1985}). The sum runs over the $1^+$ states in the intermediate nucleus and the exponent of the denominator is one for the decays to the $0^+$ states and three for the decays to the $2^+$ states. Because of the higher exponent, the cumulative matrix elements for the decays to the $2^+$ states are not only much smaller than for the decays to the $0^+$ states, but they also converge much faster as functions of the energy of the intermediate $1^+$ states. 

The third quantity in eq. (\ref{eq:2vbbhl}) is the axial-vector coupling constant $g_{\rm A}$. Its free-nucleon value determined from the neutron decay half-life is $g_{\rm A}=1.27$. However, when calculations for $2\nu\beta\beta$ decays have been carried out using the nuclear shell model and other microscopic nuclear-structure models, quenched values of $g_{\rm A}\approx1$ (or even less, see the reviews \cite{Suhonen2017,Suhonen2019}) have been consistently needed for reproducing the experimental half-lives (see also the review \cite{Ejiri2019}). In the present study, we predict branching ratios which are independent of the adopted value of $g_{\rm A}$ as long as we make the reasonable assumption that the needed effective value is the same for all the transitions. While the NMEs are not without their own nuclear-structure-related uncertainties (restricted single-particle model spaces, lacking many-body configurations and three-body forces, etc.), the quenching issue is believed to be related to the impulse approximation (see \cite{Behrens1982}) leading to the omission of the meson-exchange (or two-body) currents, shown to be quite important by Gysbers \textit{et al.} \cite{Gysbers2019}. The present shell-model calculations exploit an effective nuclear Hamiltonian where the deficiencies pertaining to nuclear structure and lack of meson-exchange currents have been (at least partially) compensated by fitting the two-body Hamiltonian to independent data in a nuclear region relevant for the present calculations. Again, the branching ratios are less sensitive to these deficiencies since they tend to cancel when taking ratios of the transition half-lives.   


The nuclear-structure calculations for the wave functions and one-body transition densities (OBTDs), needed for the evaluation of the transition half-lives, were carried out with the shell-model software NuShellX@MSU \cite{nushellx}. 

The calculations were carried out in a single-particle model space consisting of the orbitals $0f_{5/2}-1p-0g_{9/2}$ using the effective Hamiltonian JUN45 \cite{Honma2009} . The same Hamiltonian was used to successfully describe observables such as level schemes and half-lives of the neighboring nuclei $^{71}$Ge and $^{71}$Ga in \cite{Kostensalo2019}. The computational burden of carrying out the calculations in the entire model space is enormous and even using a powerful computer cluster some compromises had to be made. In any case, we managed to calculate 250 intermediate $1^+$ states in $^{76}$As using the full model space, as well as the initial state $0^+_{\rm g.s.}$ in $^{76}$Ge and the four final states of interest $0^+_{\rm g.s.}$, $0^+_2$, $2^+_1$ and $2^+_2$ in $^{76}$Se. There are a couple additional states in $^{76}$Se to which a $2\nu\beta\beta$ transition is not energetically forbidden but due to the much smaller $Q$-values and/or larger differences in angular momentum between the initial and final states, the branching ratios to these states are orders of magnitude smaller and can not be experimentally detected in the foreseeable future. The shell-model Hamiltonian managed to reproduce the low-energy spectrum of $^{76}$Se well: the ordering of the four final states was correct, with the predicted energies of the excited $2^+_1$, $0^+_2$ and $2^+_2$ states being 562 keV, 1060 keV and 1397 keV, agreeing nicely with the corresponding experimental energies of 559 keV, 1122 keV and 1216 keV \cite{Singh1995}. 

The density of the $1^+$ states in $^{76}$As is high and we ended up in a three-step shell-model calculation in order to handle these states up to excitation energies that guarantee the convergence of the $2\nu\beta\beta$ NMEs. In the first step, we managed to calculate the lowest 250 intermediate $1^+$ states in the full $0f_{5/2}-1p-0g_{9/2}$ model space, reaching 4.7 MeV in excitation energy. Since we can expect contributions from higher intermediate states to be significant, we carried out additional calculations for these states using two sequential truncations of the model space. With the first truncation, i.e. keeping the neutron orbital $0f_{5/2}$ filled, we managed to calculate also the next 250 $1^+$ states, having now access to the lowest 500 intermediate states in the second step of the calculations. In the third step we kept, in addition, the proton orbital $0g_{9/2}$ empty, which allowed us to calculate the lowest 9999 intermediate states and reach 30.6 MeV in excitation energy. These three steps were combined so that the OBTDs and excitation energies were extracted as follows: For the lowest 250 intermediate $1^+$ states we used the first-step results of the full-models-space calculation, for the next 250 $1^+$ states we used the results of the second-step truncated calculation and for the rest of the $1^+$ states we used the results of the third-step truncated calculation.
Since the density of states was highest for the first-step full calculation and lowest for the most truncated third-step calculation, we ended up with 10,266 intermediate $1^+$ states 
instead of the 9999 states of the most truncated calculation.

Since kinematics have a huge effect on the predicted half-lives, we used the available experimental information to fix the excitation energies of the lowest intermediate $1^+$ states. The calculations predict $2^-$ as the ground state in $^{76}$As, in agreement with the data. The computed excitation energies of the lowest three $1^+$ states are 185 keV, 360 keV and 519 keV, consistently by some $140-400$ keV higher than the experimental energies of 44.425(1) keV, 87 keV and 120 keV \cite{Singh1995}. There is currently no reliable experimental information on the excitation energies of the higher $1^+$ states. In order to achieve the maximal experimental input, the computed excitation energies of the mentioned $1^+$ states were shifted for all calculations so as to match the measured energies.


			\begin{figure}
	\centering	
	\includegraphics[width=0.75\textwidth]{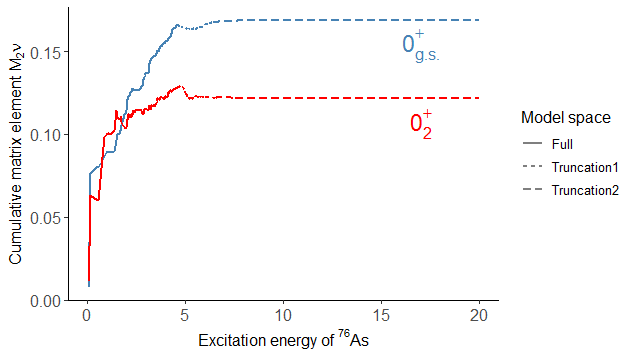}
	\caption{Cumulative $2\nu\beta\beta$ NMEs $M_{2\nu}$ for the decay of $^{76}$Ge to the $0^+_{\rm g.s.}$ and $0^+_2$ states in $^{76}$Se as functions of the excitation energy of the intermediate states in $^{76}$As. The full calculation (solid line) has been carried out in the $0f_{5/2}-1p-0g_{9/2}$ model space, for truncation 1 (dotted line) the neutron orbital $0g_{5/2}$ was kept full and for Truncation 2 (dashed line), additionally, the proton orbital $0g_{9/2}$ was kept empty.
\label{fig:0cumu}  }
	\end{figure}

			\begin{figure}
	\centering	
	\includegraphics[width=0.75\textwidth]{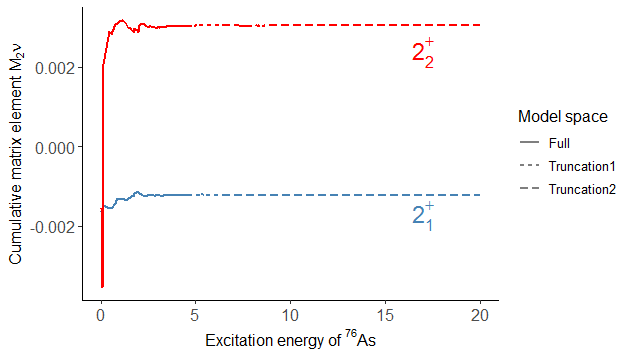}
	\caption{The same as in Fig.~\ref{fig:0cumu} for cumulative $2\nu\beta\beta$ NMEs $M_{2\nu}$ of the decay of $^{76}$Ge to the $2^+_1$ and $2^+_2$ states in $^{76}$Se. 
\label{fig:2cumu}  }
	\end{figure}
	
The cumulative nuclear matrix elements $M_{2\nu}$ for the decays to the $0^+_{\rm g.s.}$ and $0^+_2$ states are presented in Fig.~\ref{fig:0cumu} and those to the $2^+_1$ and $2^+_2$ states are given in Fig.~\ref{fig:2cumu}. The matrix elements for the decays to the $0^+$ states seem to have converged by 10 MeV in excitation energy and for the decays to the $2^+$ states by about 3 MeV in excitation energy. 

The three-step shell-model application predicts the values of the $2\nu\beta\beta$ NMEs (including the value of the axial-vector coupling $g_{\rm A}$), $\vert M_{2\nu}\vert g_{\rm A}^2$, to be 0.168$g_{\rm A}^2$, $1.2\times10^{-3}$$g_{\rm A}^2$, 0.121$g_{\rm A}^2$, and $3.1\times10^{-3}$$g_{\rm A}^2$ for the $0^+_{\rm g.s.}$, $2^+_1$, $0^+_2$ and $2^+_2$ states, respectively. When only using the first-step results, i.e. the 250 states from the full-model-space calculations, the corresponding matrix elements are 0.166$g_{\rm A}^2$, $1.2\times10^{-3}g_{\rm A}^2$, 0.128$g_{\rm A}^2$, and $3.1\times10^{-3}g_{\rm A}^2$, respectively. This means that the effects of the higher states on the final NMEs are fairly modest, with the main difference being that the NME for the decay to the $0^+_2$ state is somewhat smaller in the three-step calculation, leading to a reduction of the branching ratio to the $0^+_2$ state by 12\%  in relative terms. The second step of the calculations converges the NMEs to essentially their final values as indicated by the dotted lines in Figs.~\ref{fig:0cumu} and \ref{fig:2cumu}. The final convergence is achieved in step 3. Concerning the accuracy of the present results, one can compare the present results with those of the large-scale jun45 shell-model calculation of Caurier \textit{et al.} \cite{Caurier2012} in the same single-particle valence space. The calculation of \cite{Caurier2012} gives for the ground-state NME the value 0.170$g_{\rm A}^2$ which is almost exactly the value obtained in the present calculation. Both of these results are also in line with the result of the jun45 calculation of Brown \textit{et al.} \cite{Brown2015}.
Furthermore, one has to point out that the presently used single-particle valence space, one major shell, is rather limited and contributions from the outside of the valence space may be expected. However, these outside contributions can be, to a large extent, taken into account by the spectroscopy-fitted effective Hamiltonian used in our calculations.

\begin{figure}
	\centering	
	\includegraphics[width=0.75\textwidth]{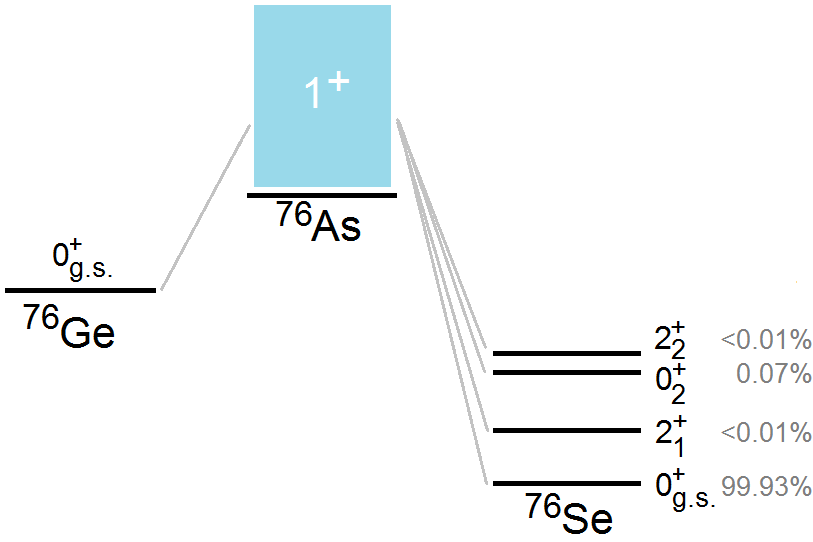}
	\caption{Scheme for the $2\nu\beta\beta$ decay of $^{76}$Ge based on the present large-scale shell-model calculations. The predicted branching ratios for the $2^+_1$ and $2^+_2$ states are 9.1$\times10^{-5}$\% and 5.0$\times10^{-7}$\%, respectively.
\label{fig:BR}  }
	\end{figure}

The computed NMEs and phase-space integrals can be used to derive the branching ratios 99.926\%, 4.4$\times10^{-5}$\%, 0.074\%, and 2.5$\times10^{-7}$\% for the transitions to the $0^+_{\rm g.s.}$, $2^+_1$, $0^+_2$, and $2^+_2$ states, respectively. These branching ratios are shown schematically in Fig.~\ref{fig:BR}. The experimental half-life of $(1.926\pm 0.094)\times10^{21}$ yr, obtained for the ground-state-to-ground-state decay \cite{GERDA2015}, can be achieved with the value $g_{\rm A}=0.80\pm0.01$ of the axial coupling, which is in line with what has been seen in many shell-model calculations, i.e., that a quenched value of the axial coupling is needed to reproduce the experimental results (see the review \cite{Suhonen2017}). Using this value of the axial coupling, the half-lives for the excited-state transitions listed in Table~\ref{tbl:NMEs} were derived. The uncertainties of the computed half-lives include only the uncertainty related to the value of $g_{\rm A}$, i.e., to the experimental half-life of the ground-state transition. This is because the uncertainties related to the matrix elements cannot be reliably estimated, though our adopted method controls for systematic under- or overestimation related to the valence space by fixing the ground-state-to-ground-state half-life using experimental data. In the same table the currently known experimental half-lives and their lower limits are shown. As can be seen, the experimental lower limit for the half-life of the transition to the $0^+_2$ state is some two orders of magnitude lower than the computed half-life. For the other transitions the experimental half-life limits are way lower than the computed half-lives.

\begin{table}
\centering
  \caption{Shell-model calculated $2\nu\beta\beta$ NMEs, phase-space integrals and half-lives  (columns 3-5) for the decay of $^{76}$Ge to the ground state and lowest three excited states (column 1) of $^{76}$Se. The phase-space integrals for the lowest three states are taken from \cite{Neacsu2016} and the fourth one has been calculated using the formulas of this reference. Experimental $Q$ values (column 2) are used in the calculations. The half-life for the ground-state transition matches the measured one \cite{GERDA2015} and the rest of the half-lives are based on the shell model calculations using $g_{\rm A}=0.80\pm0.01$, derived from the comparison of the computed half-life with the experimental one for the ground-state transition. The uncertainties include only the uncertainty related to the value of the experimental half-life of the ground-state transition. In the sixth column we list the measured lower limits for the half-lives, including the corresponding reference in the last column.} 

\begin{tabular}{ccccccc}
		\hline
 $J_f^{\pi}$ & $Q_{\beta\beta}$ (keV) &  $\vert M_{2\nu}\vert$ & $G$ (yr$^{-1}$)
&$T_{1/2}^{\beta\beta}$(th.) (yr) &  $T_{1/2}^{\beta\beta}$(exp.) (yr) & Ref. \\
  \hline
$0^+_{gs}$ & 2039 & 0.168   & $4.51\times10^{-20}$  & $(1.926\pm 0.094)\times10^{21}$ & 
$(1.926\pm 0.094)\times10^{21}$ & \cite{GERDA2015} \\
$2^+_{1}$ & 1480 &  1.2$\times10^{-3}$ & $4.0 \times10^{-22}$  & $(4.37\pm 0.20)\times10^{27}$ & $>1.1\times 10^{21}$ & \cite{Barabash1995} \\
$0^+_{2}$ & 917 & 0.121 & $6.4\times10^{-23}$ & $(2.60\pm 0.13)\times10^{24}$ & $>6.2\times 10^{21}$ & \cite{Klimenko2002}\\
$2^+_{2}$  & 823 & 3.1$\times10^{-3}$ & $3.33 \times10^{-25}$  & $(7.57\pm 0.37)\times10^{29}$ & $>1.4\times 10^{21}$ & \cite{Barabash1995} \\
\hline
\end{tabular}
\label{tbl:NMEs}

\end{table}

In Table \ref{tbl:theories} we compare the presently obtained shell-model results for the ground-state and excited-state transitions with the corresponding results of other calculations found in the literature (excluding the already mentioned two shell-model results \cite{Caurier2012,Brown2015} for the ground-state transition). For the ground-state transition the presently computed NME, 0.168, reproduces the measured half-life with 
$g_{\rm A}=0.80\pm0.01$ and the rest of the calculations with quenched values in the range 
$g_{\rm A}=0.97-1.2$. For the decay to the $2^+_{1}$ state the presently obtained result is quite close to the MCM result of \cite{Civitarese1994} and the HRPA result of \cite{Raduta2007}. Also the other results are not very far from the present result, except the HQRPA result of \cite{Stoica1996}. For the decay to the $0^+_{2}$ state the presently computed NME is quite close to the RQRPA one of \cite{Toivanen1997} and not very far from the MCM result of \cite{Civitarese1994} and the HQRPA result of \cite{Stoica1996}. Concerning the decay to the $2^+_{2}$ state, the presently obtained NME is  consistent with the MCM results of \cite{Civitarese1994,Aunola1996}. Overall, the present results for the excited-state NMEs correspond surprisingly closely to the MCM results of Civitarese \textit{et al.} \cite{Civitarese1994}.

\begin{table}
\centering
  \caption{Comparison of the presently computed $2\nu\beta\beta$ NMEs (last line) with earlier calculations for the $0^+_{\rm g.s.}$, $2^+_1$, $0^+_2$ and $2^+_2$ states (columns 1-4). Column 5 indicates the used theory, with MCM$=$Multiple Commutator Model, HQRPA$=$Higher QRPA, RQRPA$=$Renormalized QRPA, SM$=$shell model. The last column gives the reference to the calculation.}
\begin{tabular}{cccccc}
		\hline
		$\vert M_{2\nu}(0^+_{\rm g.s.})\vert$ & $\vert M_{2\nu}(2^+_{1})\vert$ & $\vert M_{2\nu}(0^+_{2})\vert$ & $\vert M_{2\nu}(2^+_{2})\vert$ & Theory & Ref. \\
		\hline
	0.074	& $1\times 10^{-3}$ & 0.363 & $3\times 10^{-3}$ & MCM & \cite{Civitarese1994}  \\
	-	& $2\times 10^{-3}$ & - & - & HQRPA & \cite{Bobyk1995}  \\
	0.100	& $3\times 10^{-3}$ & 0.838 & $3\times 10^{-3}$ & MCM & \cite{Aunola1996}  \\
	0.083	& $0.013$ & 0.056 & - & HQRPA & \cite{Stoica1996}  \\
	0.074	& $3\times 10^{-3}$ & $0.130-0.229$ & $(7-12)\times 10^{-3}$ & RQRPA & \cite{Toivanen1997}  \\
	-	& $(0.48-0.65)\times 10^{-3}$ & - & - & RQRPA & \cite{Schwieger1998}  \\
	0.113	& $0.74\times 10^{-3}$ & - & - & HRPA & \cite{Raduta2007}  \\
	0.168	& $1.2\times 10^{-3}$ & 0.121 & $3.1\times 10^{-3}$ & SM & This work  \\
\hline
\end{tabular}
\label{tbl:theories}
\end{table}


In this Letter we present the first large-scale shell-model calculations concerning the 
$2\nu\beta\beta$ decay of $^{76}$Ge to the ground state and the lowest three excited states $2_1^+$, $0_2^+$ and $2_2^+$ in $^{76}$Se. Due to the fact that the proton and neutron Fermi surfaces of the involved nuclei are roughly at the middle of the $0f_{5/2}-1p-0g_{9/2}$ major shell, adopted as the valence space of our calculations, the computational challenges are formidable even for a powerful computer cluster used in the present calculations. This is why we performed the computations of the related nuclear matrix elements by introducing a three-step procedure. The first step allowed us to extract the wave functions of the intermediate $1^+$ states in $^{76}$As in a full valence space below the low-energy limit of some 5 MeV. These states essentially determine the magnitude of the NMEs for the $2^+$ states and carry the bulk of the contributions to the NMEs of the $0^+$ states. The wave functions for the states above the low-energy limit were computed in two steps by imposing two consecutive truncations in the valence space. All the NMEs were converged very close to their final values already after the first truncation in step 2. The third step guaranteed the convergence of the NMEs, and they were subsequently used to derive branching ratios for the decays to the ground and excited states by adopting up-to-date phase-space integrals. Concerning decay transitions to these excited states, our computed branching ratios can be used as guidelines in designing future $2\nu\beta\beta$-decay experiments using a $^{76}$Ge source.

The measured half-life of the ground-state transition could be achieved by using the value $g_{\rm A}=0.80\pm0.01$ for the axial-vector coupling. Adopting this value of the axial coupling, the half-lives for the decay transitions to the ground and excited states were derived. Comparing our results for the NMEs with those of other calculations, based on higher QRPA schemes of various kinds, showed a reasonable overall agreement with most of the calculations for all NMEs. In particular, we recorded a very good agreement with the MCM results of Civitarese \textit{et al.} \cite{Civitarese1994} for the excited-state NMEs.
The step-wise shell-model calculations, introduced by us in this Letter, could possibly be used in other computationally intense $2\nu\beta\beta$-decay and/or $0\nu\beta\beta$-decay calculations in the future. An other possible way to speed up the calculations and possibly avoid model-space truncations is to use the Lanczos algorithm as described in \cite{Engel1992,Mar2003,Novario2021}.

\section*{Acknowledgements}
This work was partially supported by the Academy of Finland under the project number 318043.
We acknowledge the grants for computer resources from the Finnish Grid and Cloud Infrastructure (persistent identifier urn:nbn:fi:research-infras-2016072533)

\end{document}